\documentclass[manuscript]{aastex}

\usepackage{graphicx}
\usepackage{natbib}
\usepackage{amsmath}
\usepackage[colorlinks,urlcolor=blue,citecolor=blue,linkcolor=blue]{hyperref}

\shorttitle{Dyn. Instabilities in High-Obliquity Systems}
\shortauthors{Tamayo et al.}

\begin{document}

\title{Dynamical Instabilities in High-Obliquity Systems}

\author{D. Tamayo}
\affil{Department of Astronomy, Cornell University, Ithaca, NY 14853}
\email{dtamayo@astro.cornell.edu}

\and

\author{J. A. Burns}
\affil{Department of Astronomy \& College of Engineering, Cornell University, Ithaca, NY 14853}

\and

\author{D. P. Hamilton}
\affil{Department of Astronomy, University of Maryland, College Park, MD 20742}

\and

\author{P. D. Nicholson}
\affil{Department of Astronomy, Cornell University, Ithaca, NY 14853}

\begin{abstract}
High-inclination circumplanetary orbits that are gravitationally perturbed by the central star can undergo Kozai oscillations---large-amplitude, coupled variations in the orbital eccentricity and inclination.  We first study how this effect is modified by incorporating perturbations from the planetary oblateness.  \cite{Tremaine09} found that, for planets with obliquities $> 68.875^{\circ}$, orbits in the equilibrium local Laplace plane are unstable to eccentricity perturbations over a finite radial range, and execute large-amplitude chaotic oscillations in eccentricity and inclination.  In the hope of making that treatment more easily understandable, we analyze the problem using orbital elements, confirming this threshold obliquity.  Furthermore, we find that orbits inclined to the Laplace plane will be unstable over a broader radial range, and that such orbits can go unstable for obliquities less than $68.875^{\circ}$.  Finally, we analyze the added effects of radiation pressure, which are important for dust grains and provide a natural mechanism for particle semimajor axes to sweep via Poynting-Robertson drag through any unstable range.  We find that generally the effect persists; however, the unstable radial range is shifted and small retrograde particles can avoid the instability altogether.  We argue that this is occurs because radiation pressure modifies the equilibrium Laplace plane.
\end{abstract}

\keywords{celestial mechanics---planets and satellites: dynamical evolution and stability}

\section{INTRODUCTION}

Satellites in inclined circumplanetary orbits that are subject to gravitational perturbations from the Sun can undergo large-amplitude eccentricity oscillations through the Kozai mechanism \citep{Kozai62, Lidov62}, often with catastrophic results \cite[e.g.,][]{Carruba02}.  As discussed below, such dramatic increases of eccentricity can occur as the pericenter slows its precession, allowing the solar tugs to systematically remove angular momentum from the orbit over part of the precession cycle.  When the dominant perturbation is the Sun's gravity, this halting of the pericenter can only be achieved for highly inclined orbits ($\gtrsim 40^{\circ}$).  In this paper we consider situations where additional perturbations are also important, thereby providing new ways to slow pericenter precession and to consequently generate large eccentricities.

An important additional potential to consider is that due to the central planet's oblateness.  The study of this perturbation's effect on satellites in combination with the Sun's gravity dates back to investigations of Saturn's moon Iapetus by \cite{Laplace05}, and later, by \cite{Tisserand96}.  \cite{Allan64} subsequently generalized this analysis to an arbitrary number of perturbers.  These works were limited to circular satellite orbits, for which the motion can be expressed in terms of elementary functions.  In the general case of eccentric orbits, however, the evolution is no longer integrable.  In the special circumstance where the obliquity is zero, \cite{Kozai63}, found a class of solutions where the argument of pericenter librates around $\pm 90^{\circ}$, qualitatively similar to Kozai cycles.  \cite{Lidov74} tabulate and explore the integrable cases in the problem.  

\cite{Kudielka94} and \cite{Vashkovyak96} built on these works and discovered solutions where most or all of the orbital elements remained stationary.  However, they limited their analysis to low obliquities, applicable to the Earth-Moon system.  \cite{Tremaine09}, henceforth TTN, analyzed the full range of obliquities and found the stationary solutions for both circular and eccentric orbits.  They further provide maps of the stability of these equilibria to eccentricity and angular momentum perturbations.   Most importantly for this paper, they discovered that orbits around planets with obliquities $> 68.875^\circ$ undergo chaotic, large-amplitude oscillations in eccentricity and in inclination over the radial range from the planet where the two perturbations are comparable \cite[see Fig. 9 in][]{Tremaine09}.  For a visualization of the effect, see the orbital histories shown in the figures below.  

In this paper, we first investigate this case (including oblateness and gravitational solar perturbations) in a manner complementary to TTN.  We use orbital elements in preference to TTN's vector approach, and we derive our results from the simple condition that the pericenter be able to halt, rather than from the stability of the Laplace surface.  We thereby sacrifice some mathematical elegance and generality in order to provide a more physically intuitive picture.  In Section \ref{3d} we extend the work of TTN, which only considers orbits in the equilibrium plane, to consider the general case of orbits inclined to this Laplace plane.

Although each added perturbation greatly increases the system's complexity, in Section \ref{EM} we incorporate radiation pressure so as to be able to study the motion of dust grains, for which such forces matter \citep{Burns01}.  Most importantly, radiation forces generate Poynting-Robertson (P-R) drag. P-R drag causes an orbit's semimajor axis to decay \citep{Burns79}, allowing it to sweep through the radial range from the planet in which the eccentricity becomes unstable.  We investigate whether radiation pressure's additional effects alter this unstable radial range or are even capable of stabilizing particle orbits against the instability found by TTN.  

\section{EVOLUTION UNDER PERTURBATIONS FROM SOLAR GRAVITY AND PLANETARY OBLATENESS}

\subsection{Kozai Oscillations}
We start by reviewing the features of the Kozai mechanism that are essential to our work in order to motivate the strategy pursued in the rest of the paper.  In our context, Kozai oscillations result solely from the Sun's gravitational perturbations on a body in an inclined circumplanetary orbit.  

From a planetocentric perspective, the Sun ``orbits" the planet in the latter's orbital plane; to avoid confusion with the particle's orbital plane, we will hereafter refer to the planet's orbital plane as the ``ecliptic" (even though the latter term strictly refers to the Earth's orbital plane).  When interested in secular timescales much longer than the planet's and particle's orbital periods, one can time-average over the Sun's and particle's orbits and treat their masses as distributions smeared over their paths in the sky.  Furthermore, since a circumplanetary particle lies much closer to the planet than to the Sun, it is usually sufficient to take only the leading quadrupole term in an expansion of the solar potential in powers of $a/a_p$, where $a$ is the circumplanetary particle's semimajor axis, and $a_p$ is the planet's semimajor axis.  We point out, however, that \cite{Katz11}, \cite{Naoz11} and \cite{Lithwick11} have found that including the octupole term can introduce qualitatively different phenomena including flips from prograde to retrograde orbits, and eccentricities arbitrarily close to unity.   In the limit where the circumplanetary particle's mass is negligible, they find that the octupole correction can be ignored when $\epsilon_M \ll 1$, where 
\begin{equation}
\epsilon_M = \Bigg(\frac{a}{a_p}\Bigg)\Bigg(\frac{e_p}{1-{e_p}^2}\Bigg),
\end{equation}
and $e_p$ is the planet's orbital eccentricity.  In this paper we consider only the Sun's quadrupole potential, and our results are therefore only applicable to cases where $\epsilon_M \ll 1$.  

The secular problem truncated at quadrupole order was first analyzed by \cite{Kozai62} and \cite{Lidov62}.  We choose to work with the orbital elements ($a,e,i,\Omega, \omega$).  The equations of motion in these variables are given by \cite[][though see an erratum common to both papers in \citealt{Carruba02Err}]{Innanen97, Carruba02}:

\begin{eqnarray}
\frac{de}{dt} &=& \frac{15\epsilon_\odot n}{8} e (1 - e^2)^{1/2} {\sin}^2 i_E \: \sin 2 \omega_E \label{dedt}, \\ 
\frac{di_E}{dt} &=& -\frac{15\epsilon_\odot n}{16} e^2 (1 - e^2)^{-1/2} \sin 2 \omega_E \: \sin 2 i_E \label{didt}, \\
\frac{d\omega_E}{dt} &=& \frac{3\epsilon_\odot n}{4} (1 - e^2)^{-1/2} \Big[ 2 (1 - e^2) + 5 {\sin}^2 \omega_E(e^2 - {\sin}^2 i_E ) \Big], \label{dwdt}
\end{eqnarray}
where $n$ is the particle's mean motion, $e$ is the particle orbit's eccentricity, $\omega_E$ its argument of pericenter, and $i_E$ its inclination to the ecliptic, in which the Sun moves.  The subscript $E$ has been added to the angular quantities to emphasize that they are measured relative to the ecliptic plane.  The quantity $\epsilon_\odot$ characterizes the strength of the solar perturbation relative to the dominant planetary gravity and depends on the distance from the planet; it is given by 
\begin{equation} \label{es}
\epsilon_\odot = \frac{M_{\odot} a^3}{M_p a_p^3(1 - e_p^2)^{3/2}},
\end{equation}
where $M_p$ and $M_{\odot}$ are the planet's and Sun's masses, respectively.

The fact that the Sun has been averaged over its orbit and that the potential it creates is therefore time-independent means that energy (and thus $a$) is conserved.  Furthermore, \cite{Kozai62} realized that the problem's symmetry guaranteed the conservation of the component of angular momentum perpendicular to the ecliptic, $L_z =  \sqrt{GM_p a(1-e^2)} \cos i_E$, where $G$ is the gravitational constant.  This renders the system a one-degree-of-freedom, integrable system in the ($e$, $\omega_E$) plane, i.e., one can divide Eq. \ref{dedt} by Eq. \ref{dwdt}, eliminate $i_E$ using $L_z$, and solve for $e$ as a function of $\omega_E$.  For initial values of $\Theta = \sqrt{1-e^2} \cos i_E < 3/5$ ($i_E > 39.2^{\circ}$ for $e \ll 1$), a stable equilibrium solution exists where $e, i_E, $ and $\omega_E$ are stationary.  In this case, the phase portrait in the ($e, \omega_E$) plane consists of two types of solutions:  1) ones that trace out paths around the stationary point so that $\omega_E$ librates between minimum and maximum values, and 2) ones where $\omega_E$ circulates.  For $\Theta > 3/5$, no stationary point exists, and only circulating solutions are possible.  These behaviors can be seen in Figs. 2-8 of \cite{Kozai62} and Fig. 2 of \cite{Carruba02}.
 
More qualitatively, Eq. \ref{dedt} indicates that the pericenter's orientation within the orbital plane (given by $\omega_E$) determines whether the eccentricity grows or shrinks.  Normally $\omega_E$ circulates swiftly, as the term in brackets in Eq. \ref{dwdt} is roughly constant for  small $e$ and $i$.   This results in a small-amplitude eccentricity oscillation (due to the $\sin 2\omega_E$ term in Eq. \ref{dedt}).  However, if $d\omega_E/dt$ ever approaches zero in an orientation where $\sin 2\omega_E > 0$, the eccentricity can grow to large values.  This can occur in the Kozai problem whenever the relative inclination, $i_E$, between the particle's orbit and the distant perturber's orbit is significant.  For small eccentricities, Eq. \ref{dwdt} indicates that $d\omega_E/dt$ equals zero for some values of $\omega_E$ when $\sin^2 i_E > 2/5$ (i.e., $i_E > 39.2^\circ$).  

Large inclinations to the ecliptic therefore provide one way for the eccentricity of circumplanetary orbits to grow to large values; however, adding other perturbations may allow for additional possibilities.

\subsection{Adding Planetary Oblateness ($J_2$)} \label{addJ2}

A planet's oblateness, represented by its $J_2$ coefficient, causes pericenter precession but does not produce secular effects on an orbit's eccentricity or inclination \citep[e.g.,][]{Danby62}.  One can therefore imagine that orbital configurations may exist in which the $\omega_E$ precession from $J_2$ cancels that from the Sun, making $\omega_E$ constant and allowing the eccentricity to grow to large values according to Eq. \ref{dedt}.

The conservation of $L_z$ mentioned in the previous section is due to the quadrupole potential's azimuthal symmetry, and this causes the eccentricity and inclination evolution to be coupled \citep{Kozai62}.  But when one adds planetary oblateness, which is invariant about a different axis (the planet's spin pole), this symmetry of the classical Kozai case is destroyed.  Hence, the eccentricity and inclination become decoupled, and the optimal choice of a reference plane from which to measure all angles is no longer obvious.

\begin{figure}[!ht]
\includegraphics[width=9cm]{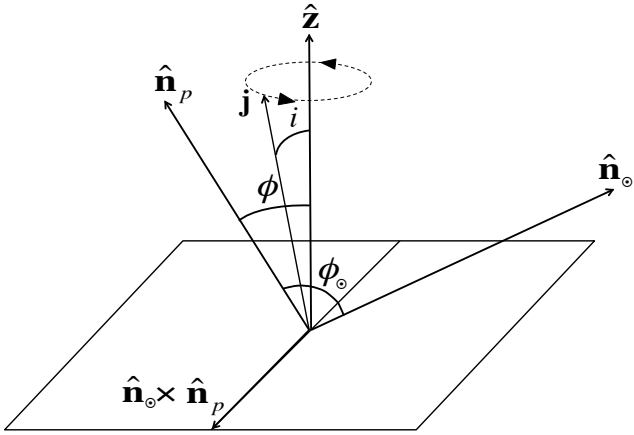}
\caption{\label{precession}  Normal to the local Laplace plane ${\bf \hat{z}}$ lies between, and is coplanar with, the planet's spin pole ${\bf \hat{n}}_p$ and the ecliptic normal ${\bf \hat{n}}_{\odot}$.  The normal to an arbitrary particle's orbit plane ${\bf j}$ will precess around ${\bf \hat{z}}$ at approximately constant inclination $i$, sweeping out a cone.  The obliquity $\phi_\odot$ is simply the angle between the vectors ${\bf \hat{n}}_p$ and ${\bf \hat{n}}_{\odot}$, and $\phi$ represents the angle between ${\bf \hat{n}}_p$ and the ${\bf \hat{z}}$ axis.  As the semimajor axis changes and the relative strengths of the Sun's and planet's perturbations vary, the Laplace plane will shift, and $\phi$ will vary.}
\end{figure}

An appropriate choice is the local Laplace plane, which lies between the planet's equatorial plane and the ecliptic.  If a particle on a circular orbit has its orbital plane align with this equilibrium plane, the torques from the Sun and $J_2$ balance so that the orbit's angular momentum vector remains fixed, and the orbital plane does not precess \citep{Allan64}.  A circular orbit {\it not} aligned with the local Laplace plane will have its orbital axis precess around the equilibrium Laplace plane axis (see Fig. \ref{precession}).  This represents a compromise between the particle orbit attempting to precess around both the planet's spin axis and the Sun's orbital axis.  

More generally, the orbit normals of eccentric orbits will wobble as they undergo their precession cycle since their changing distance from the planet means they ``sense" a range of Laplace planes.  We note that even in cases where the {\it orbital plane} itself does not precess, the {\it pericenter} may still precess within that orbital plane.  It is specifically the ability of the {\it pericenter} to halt that gives rise to large eccentricities, as argued in the first paragraph of this section.

Because the strengths of the two relevant perturbations vary differently with distance from the planet, the local Laplace plane shifts as the particle's semimajor axis varies.  Near the planet, where the torques on the orbit are predominantly caused by oblateness, the Laplace plane nearly coincides with the planet's equatorial plane.  Far from the planet, where solar torques dominate, the Laplace plane aligns closely with the ecliptic (in which the Sun ``moves").  Between these limits, the Laplace plane takes on intermediate orientations, generating a warped Laplace surface.  The Laplace plane at a given semimajor axis is the tangent plane to the Laplace surface.  The transition of the Laplace surface from the ecliptic to the equatorial plane occurs approximately at the distance where the torques from the solar tides and $J_2$ are equal.  Omitting factors of order unity, this distance is often referred to as the Laplace radius and is given by \cite{Goldreich66},
\begin{equation} \label{rl}
{r_L}^5 = J_2 {R_p}^2 {a_p}^3 (1 - e_p^2)^{3/2} \frac{M_p}{M_{\odot}},
\end{equation}
where $R_p$ and $M_p$ are the planet's radius and mass, and $M_{\odot}$, $a_p$ and $e_p$ were previously defined.  For particles orbiting at large distances from any existing inner satellites, one can treat the inner moons' effect as a further contribution to the planetary $J_2$, where (see, e.g., TTN),
\begin{equation} \label{mod}
J_2' R_p^2 \equiv J_2R_p^2 + \frac{1}{2} \sum_{i = 1}^{n} a_i^2 m_i / M_p,
\end{equation}
where $J_2'$ is the effective $J_2$, and the $a_i$ and $m_i$ are the moons' semimajor axes and masses, respectively.  Any subsequent references to a planet's $J_2$ in this paper should be understood as the effective $J_2$ that includes any inner satellites' contribution to the quadrupole potential.

An important dynamical feature for particle orbits decaying slowly compared to the precession timescale (e.g., through P-R drag) is that the orbital inclination to the local Laplace plane remains roughly constant \citep{Goldreich65}.  This means that a decaying particle orbit starting far from the primary in the planet's orbital plane (which coincides here with the local Laplace plane) will have its orbital plane follow the Laplace plane as the latter shifts on its way inward toward the planet.  

Following TTN, we neglect any variations in ${\bf \hat{n}_\odot}$, as well as the precession of ${\bf \hat{n}_p}$ due to the torques on the equatorial bulge from the Sun and other planets.  The latter timescale is generally much longer than a circumplanetary orbit's precession period, in which case it can be safely ignored \citep{Goldreich65}. 

\subsection{The Disturbing Potential}\label{distPot}

\begin{figure}[!ht]
\includegraphics[width=9cm]{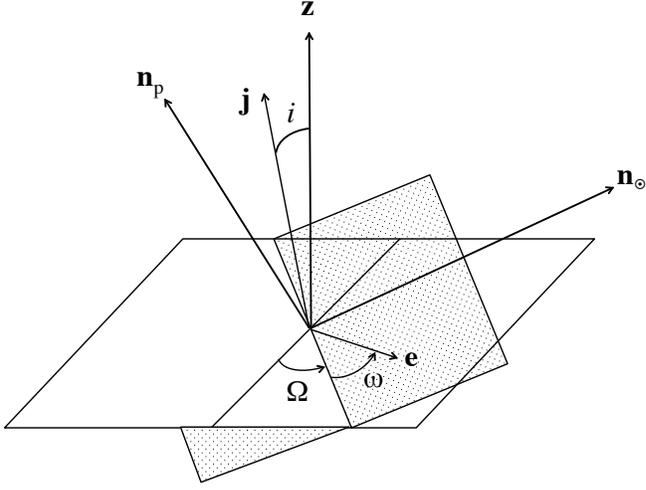}
\caption{\label{coord}  Reference plane (white) is the local Laplace plane.  ${\bf \hat{n}_p}$ is the planet's spin pole and ${\bf \hat{n}_\odot}$ the ecliptic pole.  The particle's orbital plane (shaded) is defined by its orbit normal ${\bf j}$, which can be given in terms of the orbit's inclination $i$ and longitude of the ascending node $\Omega$.  The orientation of the orbit (not shown) within the orbital plane is defined by the so-called eccentricity vector, which points toward pericenter ${\bf e}$, and is parametrized by the argument of pericenter $\omega$, measured along the shaded orbital plane, from the line of ascending node.  We choose to measure the longitude of the ascending node $\Omega$ from the direction defined by ${\bf \hat{n}_\odot \times \hat{n}_p}$.  Note that ${\bf j}$ and ${\bf e}$ are not unit vectors.}
\end{figure}

We now derive the disturbing potential using orbital elements and TTN's notation.  The obliquity $\phi_\odot$ and the variable $\phi$ are defined in Fig. \ref{precession}, whereas the remaining quantities are shown in Fig. \ref{coord}. 

The eccentricity vector ${\bf e}$ points toward pericenter and has a magnitude given by the orbit's eccentricity.  The vector ${\bf j}$ lies along the orbit normal and has a magnitude chosen as $\sqrt{1 - e^2}$, so that ${\bf j} \cdot {\bf e} = 0$ and $j^2 + e^2 = 1$.  We choose to measure the longitude of the ascending node $\Omega$ from the direction along ${\bf \hat{n}_\odot \times \hat{n}_p}$.

The perturbing potentials (averaged over both the particle and solar orbits) are (see Eq. 12 in TTN):
\begin{eqnarray}\label{psis}
\Psi_p &=& \frac{\epsilon_p}{4(1 - e^2)^{5/2}}[1 - e^2 - 3({\bf j} \cdot {\bf \hat{n}_p})^2]  \nonumber \\
\Psi_{\odot} &=& \frac{ 3\epsilon_\odot}{8} [5({\bf e} \cdot {\bf \hat{n}}_{\odot})^2 - ({\bf j} \cdot {\bf \hat{n}}_{\odot})^2 - 2e^2], \label{psi}
\end{eqnarray}
where $\Psi_p$ and $\Psi_{\odot}$ represent potentials non-dimensionalized by the factor $GM_p / a$.  The quantity $\epsilon_\odot$ was defined in Eq. \ref{es} and $\epsilon_p$ is given by
\begin{equation}  \label{ep}
\epsilon_p = \frac{J_2 R_p^2}{a^2}.
\end{equation}
Since our aim is to approach the problem using orbital elements, we employ them to rewrite the three scalar products in Eqs. \ref{psi}.  This is perhaps easiest to do by first writing ${\bf j}$, ${\bf \hat{n}}_p$, ${\bf \hat{n}}_{\odot}$ and ${\bf e}$ in terms of their Cartesian components.  This process yields
\begin{eqnarray}
{\bf j} \cdot {\bf \hat{n}}_p &=& (1 - e^2)^{1/2} (\cos{ \phi} \cos i + \sin \phi \sin i \cos \omega) \nonumber \\
{\bf j} \cdot {\bf \hat{n}}_{\odot} &=& (1 - e^2)^{1/2} [\cos{(\phi_{\odot} - \phi)} \cos i - \sin(\phi_{\odot} - \phi) \sin i \cos \Omega] \\
{\bf e} \cdot {\bf \hat{n}}_{\odot} &=& e [ \sin(\phi_{\odot} - \phi)(\cos i \cos \Omega \sin \omega + \cos \omega \sin \Omega) + \cos(\phi_{\odot} - \phi)\sin i \sin \omega]. \nonumber 
\end{eqnarray}
Substituting into Eqs. \ref{psis}, we obtain the dimensionless disturbing function $R = - (\Psi_p + \Psi_{\odot})$, 
\begin{eqnarray}\label{R}
R = - \frac{3\epsilon_\odot}{8}\Bigg\{&&\frac{2}{3(1 - e^2)^{3/2}} \Big(\frac{r_L}{a}\Big)^5 [1 - 3(\cos\phi \cos i + \sin \phi \sin i \cos \Omega)^2] + \nonumber \\  
&&5e^2[\sin(\phi_{\odot} - \phi) (\cos i \cos \Omega \sin \omega + \cos \omega \sin \Omega) + \cos(\phi_{\odot} - \phi) \sin i \sin \omega]^2 - \nonumber \\ 
&&(1 - e^2)(\cos(\phi_{\odot} - \phi) \cos i - \sin(\phi_{\odot} - \phi) \sin i \cos \Omega)^2 - 2e^2 \Bigg\},
\end{eqnarray}
where we have written the potential only in terms of $\epsilon_\odot$ in order to explicitly bring out the dependence on the semimajor axis $a$ relative to the Laplace radius $r_L$ (note that from Eqs. \ref{rl}, \ref{ep} and \ref{es}, $\epsilon_p  / \epsilon_\odot = (r_L/a)^5$).  

Implicit in Eq. \ref{R} is a relation between $a$, $\phi$ and $\phi_{\odot}$, since the semimajor axis sets the location of the local Laplace plane and therefore of the $\bf{z}$ axis (see Fig. \ref{coord}).  Changing $a$ therefore alters $\phi$.  The transcendental equation that connects these quantities is \citep{Tremaine09}
\begin{equation} \label{trans}
\tan 2\phi = \frac{\sin 2 \phi_{\odot}}{\cos 2\phi_{\odot} + 2\Big(r_L/a\Big)^5}.
\end{equation}
Equation \ref{trans} has four solutions in a $2\pi$ interval: $\phi$, $\phi + \pi$, and $\phi \pm \pi/2$.  The solution corresponding to the classical Laplace equilibrium has the property $\phi \rightarrow \phi_{\odot}$ for $a \gg r_L$.  Orbits locally aligned with this surface are stable to small perturbations in their orientation.  The other two solutions $90^{\circ}$ away are always unstable:  if the orbit plane is displaced slightly from either of these directions, it will precess about the stable solutions; we therefore do not discuss them further (see TTN).  

Eq. \ref{R} is equivalent to the disturbing function in Eq. 1.3 of \cite{Lidov74}, which is instead referenced to the ecliptic.  The two are simply related by a rotation by $\phi_\odot - \phi$ about the x-axis, where $\phi$ is given by Eq. \ref{trans}.  

\subsection{Dynamics in the Laplace Plane}

While orbits in the classical Laplace plane are always stable against perturbations to their orbit normal, they are not always stable to small changes in their eccentricity.  We now find the radial distance at which the Laplace plane becomes unstable to such eccentricity perturbations by considering the limit $i \rightarrow 0$.  As the orbital plane approaches the Laplace plane, however, $\Omega$ becomes ill-defined.  One therefore typically switches to the variable $\varpi = \Omega + \omega$, which smoothly approaches the angle between ${\bf \hat{n}_\odot \times \hat{n}_p}$ and pericenter (see Fig. \ref{coord}).

Before continuing, we wish to relate the angular variables to $\omega_E$, since Eq. \ref{dedt} shows that it is specifically the halting of $\omega_E$ that can create large eccentricities.  Careful inspection of Fig. \ref{coord} shows that orbits in the Laplace plane (the white reference plane) must intersect the ecliptic plane (not shown but perpendicular to ${\bf \hat{n}_\odot}$) along the vector ${\bf \hat{n}_\odot \times \hat{n}_p}$.  If we make the same choice of ${\bf \hat{n}_\odot \times \hat{n}_p}$ as the arbitrary reference direction in the ecliptic plane, this means that orbits in the Laplace plane always satisfy $\Omega_E = 0$.  The angle $\omega_E$ is then just the angle from ${\bf \hat{n}_\odot \times \hat{n}_p}$ to pericenter, or $\varpi$.  Therefore, for orbits in the Laplace plane, we set $i = 0$ in Eq. \ref{R} and, writing $\Omega + \omega = \omega_E$, we obtain  
\begin{equation} \label{Rsimple}
R = - \frac{3\epsilon_\odot}{8}\Bigg\{\frac{2}{3 (1 - e^2)^{3/2}} \Big(\frac{r_L}{a}\Big)^5\Big(1 - 3 \cos^2 \phi \Big) + 5e^2 \sin^2(\phi_{\odot} - \phi) \sin^2 \omega_E - (1 - e^2) \cos^2(\phi_{\odot} - \phi) - 2e^2\Bigg\}.
\end{equation}

We can now employ Lagrange's planetary equations to find the orbital elements' time evolution.  The equation for the eccentricity is \cite[cf.][p. 251, noting that we have non-dimensionalized R and that over-dots denote time derivatives]{Murray99},
\begin{equation} \label{edoteq}
\frac{\dot{e}}{n} = -\frac{(1 - e^2)^{1/2}}{e} \frac{\partial R}{\partial \varpi} =  -\frac{(1 - e^2)^{1/2}}{e} \frac{\partial R}{\partial \omega_E},
\end{equation}
where we have ignored a term involving the mean longitude at epoch that disappears in our orbit-averaged equations.
Plugging Eq. \ref{Rsimple} into this equation,
\begin{equation} \label{edot}
\frac{\dot{e}}{n} = \frac{15}{8} \epsilon_\odot e (1 - e^2)^{1/2} \sin^2 (\phi_{\odot} - \phi) \sin 2 \omega_E.
\end{equation}
Since we are considering orbits in the Laplace plane, {\bf j} lies along ${\bf \hat{z}}$ in Fig. \ref{precession}, and since $i_E$ refers to the angle between ${\bf j}$ and the ecliptic axis ${\bf \hat{n}_\odot}$, $i_E = \phi_\odot - \phi$.  Hence Eq. \ref{edot} matches the classical Kozai result (Eq. \ref{dedt}).  This is what one would expect as the planet's oblateness does not contribute secularly to the eccentricity evolution \citep[e.g.,][]{Danby62}.   

The precession of pericenter {\it is} altered by the planet's oblateness, however.  The appropriate equation is \cite[cf.][p. 251]{Murray99},
\begin{equation}\label{wdoteq}
\frac{\dot{\omega}_E}{n} = \frac{\dot{\varpi}}{n} = \frac{(1 - e^2)^{1/2}}{e} \frac{\partial R}{\partial e},
\end{equation}
where we have omitted a term $\propto \partial R / \partial i$ that vanishes in the limit $i \rightarrow 0$.  Again substituting for $R$ from Eq. \ref{Rsimple},
\begin{equation} \label{second}
\frac{\dot{\omega}_E}{n} = \frac{3\epsilon_\odot(1 - e^2)^{1/2}}{4} \Big\{ 2 - 5 \sin^2 (\phi_{\odot} - \phi) \sin^2 \omega_E - \cos^2 (\phi_{\odot} - \phi) - \frac{1}{(1 - e^2)^{5/2}}\Big(\frac{r_L}{a} \Big)^5 [ 1 - 3 \cos^2 \phi ]\Big\}.
\end{equation}
In the limit $a \ll r_L$, $\phi \rightarrow 0$, and $\dot{\omega}_E \rightarrow (3/2)\epsilon_p(1-e^2)^{-2}$ (where, once again from below Eq. \ref{R}, $(r_L/a)^5 = \epsilon_p/\epsilon_\odot$).  This matches the rate of precession $\dot{\varpi}_p$ for the longitude of pericenter relative to the planet's equatorial plane due to oblateness \citep{Danby62}.  In the limit $a \gg r_L$, $\phi_\odot \rightarrow \phi$ (Eq. \ref{trans}) and $\dot{\omega}_E \rightarrow (3/4) \epsilon_\odot (1-e^2)^{1/2}$.  Since we have restricted ourselves to orbits in the Laplace plane, this does not agree with Eq. \ref{dwdt}; rather it provides the rate one would obtain solely from solar perturbations after imposing the Laplace-plane condition that $\Omega_E = 0$ (see the paragraph preceding Eq. \ref{Rsimple}).  

Recall that inducing large eccentricity amplitudes relies on the pericenter precession rate approaching zero, i.e., keeping $\omega_E$ constant (cf. Eq. \ref{dedt}).  In the above two limiting cases, $\dot{\omega}_E > 0$, so if $\dot{\omega}_E$ is to cross through zero, it must do so at intermediate semimajor axes.  To find the least stringent condition for the pericenter to lock, one can pick the most unstable configuration (i.e., the orientation that generates the largest negative terms).  As in the Kozai case, this corresponds to $\sin^2 \omega_E = 1$, i.e., $\omega_E = \pm 90^{\circ}$.  Setting $\dot{\omega}_E = 0$ in Eq. \ref{second} with $\omega_E = \pm 90^{\circ}$ yields, to first order in $e$,
\begin{equation} \label{cond}
\frac{3\epsilon_\odot}{4} \Big\{ 1 - 4 \sin^2 (\phi_{\odot} - \phi) - \Big(\frac{r_L}{a} \Big)^5 [ 1 - 3 \cos^2 \phi ]\Big\} = 0.
\end{equation}
One cannot analytically find a solution for $a$ since $\phi$, $\phi_{\odot}$ and $a$ are all related through the transcendental relation in Eq. \ref{trans}.  But one can see that $4\sin^2(\phi_{\odot} - \phi)$ will only be large for $\phi$ far from $\phi_{\odot}$, and the last term (from $J_2$) is only negative for $\phi \gtrsim 55^{\circ}$.  Since $\phi$ is bounded to be between zero and the obliquity $\phi_{\odot}$ (see Fig. \ref{coord}), this suggests that high obliquities are required for $\dot{\omega}_E$ to drop below zero.  It also means that the roots of Eq. \ref{cond} (if they exist) should be close to $r_L$, since this is where the Laplace plane transitions and is the only situation where $\phi_\odot - \phi$ and $r_L/a$ are simultaneously appreciable.

When Eq. \ref{cond} is solved numerically for various $\phi_{\odot}$, no solution appears for $\phi_{\odot} <  68.875^{\circ}$.  Below this obliquity, orbits are therefore always stable.  Beyond this threshold obliquity, however, the pericenter can halt for a range in $a$, thereby generating large-amplitude eccentricity oscillations; our value for the critical $\phi_{\odot}$ agrees with that derived differently by TTN.  

\subsection{Uranus:  A case study}

\begin{figure}[!ht]
\includegraphics[width=12cm]{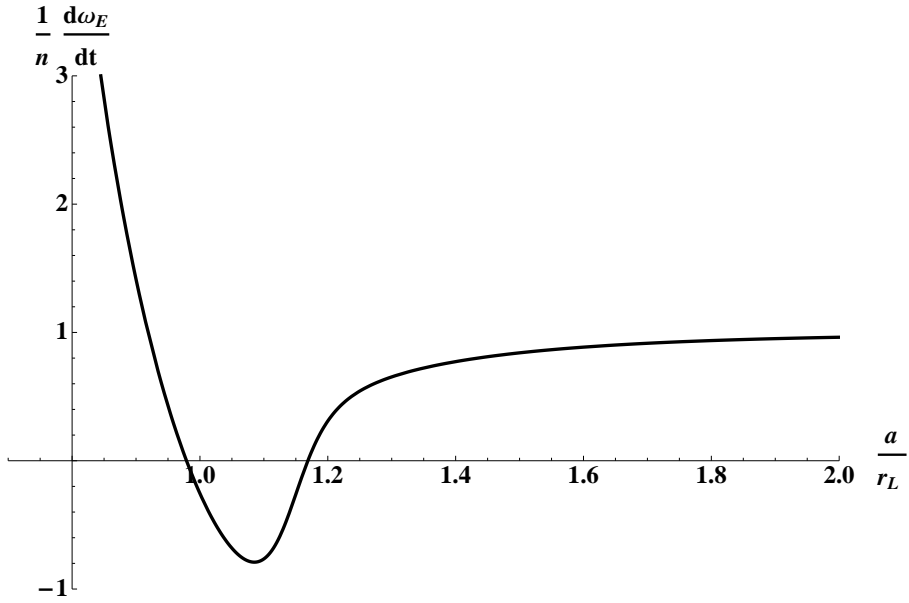}
\caption{\label{po}  For low eccentricities, the minimum $\dot{\omega}_E/n$ (at $\omega_E = \pm 90^{\circ}$ in Eq. \ref{second}) as a function of semimajor axis.  The semimajor axis is in units of the Laplace radius $r_L \approx 64 R_p$ for Uranus (Eq. \ref{rl}).  The non-dimensionalized precession rate is expressed as a fraction of the rate for $a \gg r_L$ $(\dot{\omega}_E/n = 3\epsilon_\odot/4)$.  In the radial range where $(\dot{\omega}_E/n)_{min} <$ 0, $\dot{\omega}_E/n$ will cross through 0 for certain values of $\omega_E$.  In this radial range, the Laplace plane is unstable to eccentricity perturbations. }
\end{figure}

Uranus is a solar system example with an extreme obliquity ($\phi_{\odot} \approx 98^{\circ}$) beyond the threshold value of $68.875^{\circ}$.  Hence, circumplanetary particles within a certain semimajor axis range will generate large eccentricity values.  This unstable range is depicted in Fig. \ref{po}, which prescribes the minimum non-dimensionalized precession rate $\dot{\omega}_E/n$ as given by Eq. \ref{second} for low eccentricities and $\omega_E = \pm 90^{\circ}$, plotted vs. semi-major axis (using Eq. \ref{trans} to solve for $\phi$).  

Fig. \ref{po} shows that a circular orbit lying in the classical Laplace plane will be unstable in the approximate range $0.93r_L < a < 1.17 r_L$.  In the case of Uranus, the effective $J_2$ including the contribution of the inner satellites is approximately $0.019$ (Eq. \ref{mod}), $r_L \approx 64 R_p$, and the unstable range translates to $59.5 R_p < a < 74.9 R_p$.   

\begin{figure}[!ht]
\includegraphics[width=12cm]{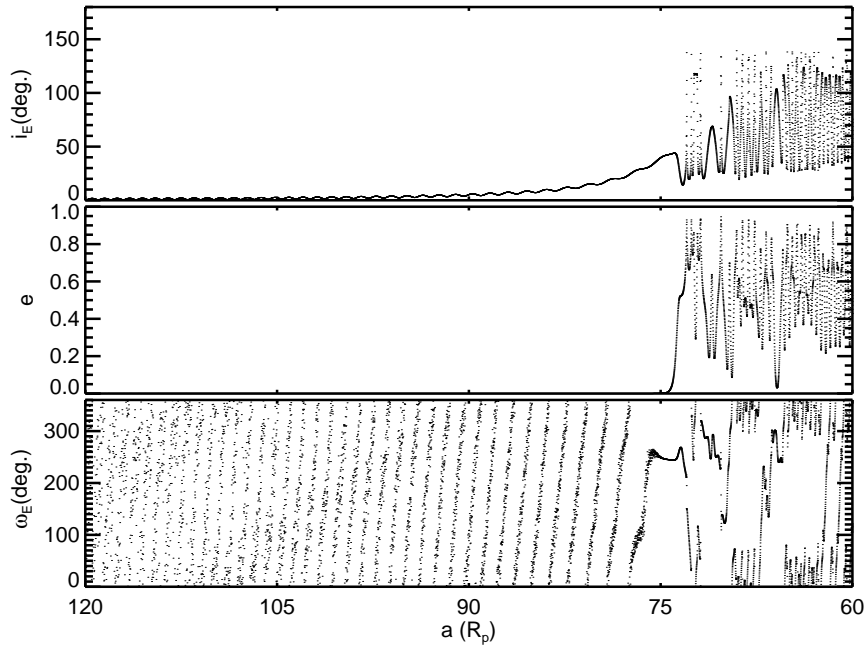}
\caption{\label{circ} Numerical integration of an initially nearly circular orbit started in the ecliptic at $120 R_p$ and slowly brought inward.  The top panel plots inclination referenced to the ecliptic, so initially $i_E = 0$.  The middle plot displays the eccentricity history, and the bottom plot shows the evolution of $\omega_E$.  The eccentricity and inclination become unstable when the semimajor axis reaches $\approx 74.9 R_p$.  Note also that this is the point where $\omega_E$ remains constant, near the most unstable orientation $\omega_E = 270^{\circ}$.}
\end{figure}

Fig. \ref{circ} displays a numerical integration of a nearly circular orbit (initial eccentricity of $10^{-6}$) started far from the planet in the ecliptic (coincident with the local Laplace plane).  The particle is then slowly brought inward according to $a=a_0e^{-t/\tau}$, where $\tau = 2.5 \text{Myr}$.  This interval is much longer than the secular timescale on which the orbit evolves of $\sim0.01 \text{Myr}$.  The functional form for the semimajor-axis decay was chosen to match that for P-R drag \citep{Burns79}; this is simply for consistency with Sec. \ref{rad} where we consider small particles that are subject to this dissipative force.  Uranus' orbit is taken as circular, and the whole third body effect of the Sun is included.  As our later integrations will include radiation pressure, we used the well-established dust integrator dI for all our numerical simulations \citep{Hamilton93, Hamilton96, Hamilton96Mars, Hamilton08, Tamayo11, Jontof12, Jontof12b}.

Since the particle starts in the ecliptic, the inclination relative to the ecliptic $i_E$ begins at zero.  As the orbit approaches the Laplace radius (64 $R_p$), the inclination follows the local Laplace plane toward Uranus' equatorial plane.  However, both the eccentricity and inclination become unstable immediately upon entering the unstable range at $\approx 74.9 R_p$.  One can see this corresponds to the point where $\omega_E$ (bottom panel) stops precessing (at the maximally unstable orientation of $270^{\circ}$).  The reason the eccentricity does not grow the first time $\dot{\omega}_E$ drops to zero at $a \approx 77 R_p$ is that $\omega_E$ is just above $270^{\circ}$, where according to Eq. \ref{dedt}, the eccentricity shrinks.  The second time $\dot{\omega}_E < 270^{\circ}$, so $\dot{e} > 0$.  Once the eccentricity grows, $e$ eventually becomes large enough that the second-order eccentricity contribution to the last term of Eq. \ref{second} becomes important and causes precession to resume, i.e., the particle gets close enough to Uranus at pericenter that $J_2$ re-initiates precession.

As mentioned at the beginning of Sec. \ref{addJ2}, this behavior differs from that of Kozai cycles.  The oscillations are not regular and the eccentricity and inclination are not coupled.  In the Kozai case this coupling was due to the conservation of angular momentum along the ecliptic axis.  The planet's oblateness spoils this symmetry from the Kozai problem because it allows the particle to exchange substantial angular momentum with the planet at pericenter along that previously conserved direction.  Note that when the eccentricity becomes large and the particles at pericenter approach the inner moons, the approximation used in our calculations and simulations of treating the inner satellites as a contribution to the planet's $J_2$ is no longer an appropriate assumption.  However, one would expect collisions to remove particles shortly after their orbits cross those of these large satellites.

We also point out that even though the classical Laplace plane is defined only for circular orbits, eccentric Laplace equilibria also exist (TTN).  The circular and eccentric equilibria bifurcate when the circular solution becomes unstable, and a decaying particle-orbit can transfer onto the eccentric-equilibrium track.  However, TTN find that the eccentric equilibrium becomes unstable almost immediately upon bifurcating from the circular solution (see Fig. 6 of TTN).  This is especially true for particles starting far from the planet and evolving inward through the unstable region (rather than starting close and evolving outward).  The point at which the classical Laplace plane becomes unstable (found above) is therefore a good proxy for when a circular orbit originally in the Laplace plane destabilizes.

\section{DYNAMICS IN THREE DIMENSIONS} \label{3d}

By restricting the above discussion to orbits that lie in the local Laplace plane, we reduced the dimensionality of the problem to two dimensions.  We now address the general situation where orbits are inclined to the local Laplace plane.  Then the problem is inherently three-dimensional, and the enlarged phase space makes it difficult to provide detailed general results.  Accordingly, we do not pursue a complete analytical theory and instead limit ourselves to a qualitative description of orbital behavior based on our understanding derived from the above analysis as well as various numerical integrations.  There are some integrable cases considered by \cite{Lidov74}; however, in the case of interest with finite eccentricity and $a \sim r_L$, these only apply to obliquities of zero and ninety degrees, or to polar orbits with the orbital axis pointing along the intersection between the planet's orbital and equatorial planes.

We can first gain some insight by investigating the equations of motion relative to the ecliptic plane.  Since $J_2$ perturbations have no secular effect on $e$, the eccentricity evolution depends only on solar perturbations; it is therefore given simply by Eq. \ref{dedt}.   In Eqs. \ref{didt} and \ref{dwdt} for $di/dt$ and $d\omega_E/dt$, one would have to add the complicated effect of $J_2$ referenced to the ecliptic plane.  This is obtained by taking the $J_2$ contribution to $R$ in Eq. \ref{R}, i.e. the term involving $(r_L/a)^5$, setting $\phi = \phi_{\odot}$, appending a subscript $E$ to all angular variables, and applying \cite[cf.][recalling that our disturbing potential is non-dimensional]{Danby62}
\begin{eqnarray}
\frac{\dot{i_E}}{n} &=& -\frac{1}{\sqrt{1 - e^2}} \Big\{ \csc \: i_E \frac{\partial R}{\partial \Omega_E} - \cot\:i_E \frac{\partial R}{\partial \omega_E}\Big \} \\
\frac{\dot{\omega}_E}{n} &=& \frac{\sqrt{1 - e^2}}{e} \frac{\partial R}{\partial e} - \frac{1}{\sqrt{1 - e^2}} \cot\:i_E \frac{\partial R}{\partial i_E}. \nonumber
\end{eqnarray}
The resulting equations are complex and difficult to pursue analytically.  One can, however, gain insight from investigating the effect of a non-zero inclination on the solar perturbations that dominate the particle's early evolution far from the planet, before the $J_2$ terms become important.    

From Eq. \ref{dwdt}, a larger inclination acts to lower $d\omega_E/dt$, bringing the orbit closer to instability.  This leads to Kozai oscillations for sufficiently large $i_E$ in this limit that ignores $J_2$.  In this sense, the previous section's situation where a particle begins far from the planet in the Laplace plane (where $i_E = 0$) furnishes the best-case scenario for stability; $d\omega_E/dt$ would have to be substantially decreased by $J_2$ in order for $d\omega_E/dt$ to drop to zero.  One should therefore expect that, if a circular orbit starting in the Laplace plane becomes unstable, any orbit initially inclined to the ecliptic will also destabilize.  Furthermore, inclined orbits should become unstable earlier during their inward evolution than their uninclined counterparts would.

\begin{figure}[!ht]
\includegraphics[width=12cm]{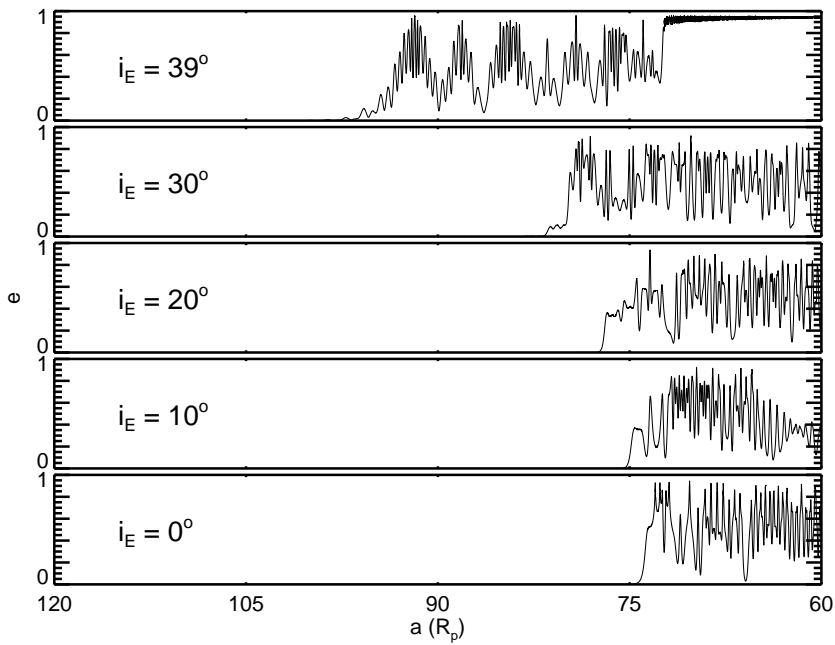}
\caption{\label{inccomp}  Orbital eccentricity histories for particles begun far from Uranus ($120 R_p$) with $e=10^{-6}$ at varying inclinations to the ecliptic.  Like Fig. \ref{circ}, the semimajor axis is brought inward according to $a=a_0e^{-t/\tau}$, with $\tau=2.5$Myr.  The figure plots eccentricity vs. semimajor axis, where constant offsets have been added to the eccentricities to separate the different plots.  Higher-inclination orbits are inherently less stable and undergo large-amplitude eccentricity oscillations sooner in their inward evolution.}
\end{figure}

Numerical integrations support this assertion.  Figure \ref{inccomp} shows the evolution of nearly circular orbits that are slowly evolved inward after beginning far from Uranus at various initial inclinations to the ecliptic.  One sees that more inclined orbits become unstable earlier in their inward migration.  We point out, however, that the orderly progression in Fig. \ref{inccomp} results from starting all the integrations with the same initial conditions (other than the inclination).  In the general 3-D case, all the orbital elements affect the dynamics.  Altering the initial conditions changes the phase in the elements' evolution at which they enter the unstable range in $a$ that we found in the 2-D case; this can change the semimajor axis at which the eccentricity grows by $\sim10\%$.  It nevertheless remains true that for a given set of initial conditions, increasing the inclination destabilizes the orbit earlier in its inward evolution.

Changing the initial $e$, however, does not have a strong effect on the location at which the pericenter halts and the orbit undergoes large-amplitude eccentricity oscillations.  This can be seen from Eq. \ref{second}.  A non-zero $e$ enhances the $J_2$ contribution, pushing outward the location at which $e$ grows slightly; however, this term's steep dependence on semimajor axis of $(r_L/a)^5$ allows a small change in $a$ to accommodate a large initial eccentricity.  The edge of the unstable region therefore shifts by less than a few percent for $e \lesssim 0.3$.

A more complete investigation of inclined orbits is beyond the scope of this paper.  We note, however, that the threshold obliquity of $68.875^{\circ}$ found by TTN, and derivable from Eq. \ref{cond}, applies to orbits in the Laplace plane, i.e., the most stable configuration.  For orbits initially inclined to the Laplace plane, the threshold obliquity would be lower.  An inward-evolving object with an initial inclination close to the threshold value for Kozai oscillations ($\approx 39.2^{\circ}$ for low eccentricities) could undergo large-amplitude eccentricity oscillations in systems with more modest obliquities.  We have verified this, finding that for a hypothetical planet with obliquity $\phi_\odot = 60^{\circ}$, orbits begun with $e=0$ and $i_E \gtrsim 7^{\circ}$ undergo eccentricity oscillations in the transition region.  The maximum eccentricity attained increases with initial inclination, varying from $e_{max} \approx 0.35$ for initial $i_E = 10^{\circ}$ to $e_{max} \approx 0.95$ for $i_E = 35^{\circ}$.  Even at Saturn ($\phi_\odot \approx 27^{\circ}$), initially circular orbits started with $i_E = 35^{\circ}$ undergo oscillations with a maximum eccentricity of $\approx 0.2$.  

\section{THE EFFECTS OF NON-GRAVITATIONAL FORCES} \label{EM}
The mechanism discussed in this paper occurs when the effects of planetary oblateness and solar gravity balance at $a \sim r_L$ (Eq. \ref{rl}) so as to halt pericenter precession.  It is therefore appropriate to consider whether additional perturbations might instead keep the pericenter moving, thereby stabilizing the orbit.  While the previous discussion applies to objects of arbitrary size and mass, we now consider radiation forces, which are most important for small dust grains.  

Radiation forces are of particular interest because they provide a natural mechanism (P-R drag) for the semimajor axis of dust-grain orbits to decay in toward the planet and reach the unstable range in $a$ \citep{Burns79}.  Two further effects that are generally important for dust grains are direct solar radiation pressure \citep{Burns79} and electromagnetic forces due to the planetary magnetic field \citep{Hamilton93, Hamilton96, Burns01}.

At all the solar system's planets except Jupiter, the instability occurs beyond the magnetopause, rendering perturbations from the planetary magnetic field irrelevant.  Furthermore, radiation pressure (discussed below) can remove small particles by pumping their eccentricities close to unity.  Particles then either crash into the primary or escape the system entirely \cite[see][]{Hamilton92}.  For dust grains starting far from the planet ($\sim 200 R_p$), only particles larger than roughly a few microns in radius survive (see below).  For particles of this size and larger, even inside the magnetosphere, the planetary magnetic field is not important \cite[cf. Fig. 11 in][]{Burns01}; we therefore ignore it.  

\subsection{Radiation Pressure} \label{radpsec}
Solar radiation pressure, however, can have powerful effects.  This perturbation has been extensively studied, usually by approximating the planetary orbit as circular and averaging over the particle's orbit, which generally changes much faster than the planet's orbital period \citep{Burns79, Hamilton93, Juhasz95}.  \cite{Mignard84} found an exact solution under these assumptions in a frame rotating with the Sun, employing several changes of variables; unfortunately, the inverse transformations to the orbital elements that we have utilized are complicated.  We therefore choose to instead work in the same inertial system we employed above and to find approximate equations sufficient for our needs.  

Upon averaging over the particle's orbit, two fundamental timescales remain.  The first is the secular rate at which the orbital elements change ($\sim n F_{rad} / F_g$, where $n$ is the particle's mean motion and $F_{rad}$ and $F_g$ are the radiation pressure and planet's gravitational forces, respectively); the second is simply the Sun's mean motion about the planet $n_\odot$.  The dynamics are set through their ratio $Z \equiv (3 n F_{rad}) / (2n_\odot F_g)$, where the factor of $3/2$ results from the exact form of the equations of motion \citep{Burns79}.  Note that since we will later be interested in higher-order eccentricity terms, we have removed the changing factor of $\sqrt{1-e^2}$ from the definition of \cite{Burns79} so that $Z$ is constant (at a given semimajor axis).  One can express $Z$ as
\begin{equation} \label{Z}
Z = 0.86 Q_{pr} \Bigg(\frac{1\text{ g cm}^{-3}}{\rho}\Bigg) {\Bigg(\frac{1\: \mu\text{m}}{s}\Bigg)\Bigg(\frac{M_\odot}{M_p}\Bigg)^{1/2} \Bigg(\frac{a}{a_p}\Bigg)}^{1/2},
\end{equation} 
where $Q_{pr}$ is the radiation pressure coefficient averaged over the central star's spectrum, $\rho$ is the particle density and $s$ is the particle radius.  Note that smaller particles, with larger surface-area-to-volume ratios, are more affected by radiation pressure (i.e., have higher Z).  However, once particles shrink below the scale of the star's peak emission wavelength, they lose the ability to couple to the radiation field and $Q_{pr}$ drops to zero; this occurs at $\sim 0.1 \:\mu$m for solar-type stars.  

If $Z$ approaches one, radiation pressure pumps a particle's orbital eccentricity to unity, most often resulting in collision with the planet.  This provides a minimum particle size to consider.  However, because $Z$ increases with $a$, this limit would vary with initial location from the planet.  This is because the importance of radiation pressure relative to the dominant planetary gravitational field increases the farther out one orbits in the primary's gravitational well.  For a more detailed analysis of this threshold and the ultimate fate of these grains, see \cite{Hamilton92}.  

\begin{figure}[!ht]
\includegraphics[width=12cm]{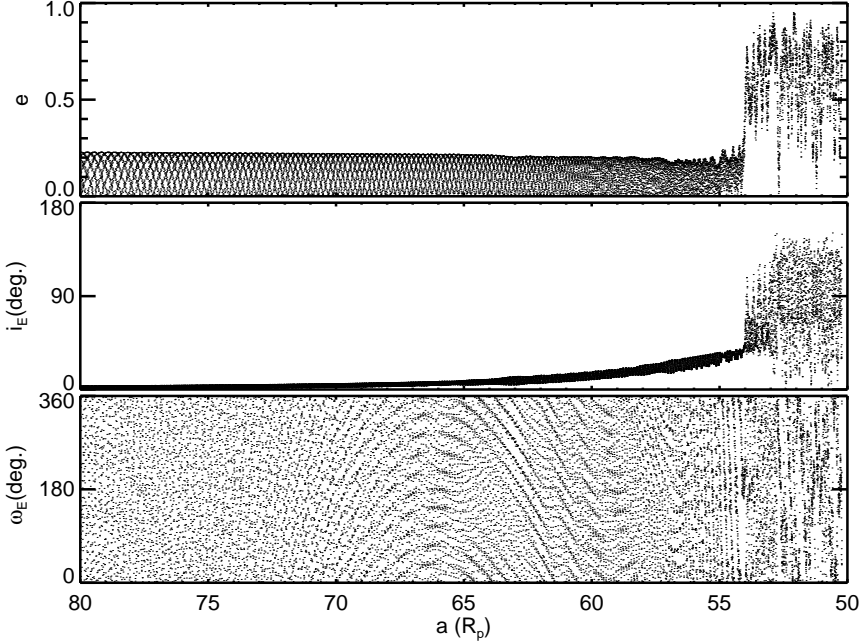}
\caption{\label{rad}  Orbital integration of a $Z=0.1$ particle begun at $80 R_p$ from Uranus with $e=10^{-6}$ and $i_E \approx 0.06^{\circ}$.  Radiation pressure has caused the location of the Laplace plane's transition to shift inward from $a \approx 75 R_p$ to $a \approx 55 R_p$ (cf. Fig. \ref{circ}).}
\end{figure}

The inclusion of radiation pressure introduces high-frequency variations to all the orbital elements at the Sun's orbital rate about the planet, $n_\odot$ \citep{Burns79}.  Since these rates are much faster than the precession rates due to $J_2$ and solar tides, one can average over these fast solar oscillations.  As shown in the next section, to first order in the eccentricity, no secular change in $\omega_E$ occurs.  For orbits in the Laplace plane with low eccentricities, Eq. \ref{cond} therefore remains the {\it averaged} condition for the pericenter to halt (at $\omega_E = \pm 90^{\circ}$).  Thus, as we argued following Eq. \ref{cond}, $\dot{\omega}_E$ will still only cross through zero at approximately the semimajor axis where the Laplace plane transitions from the ecliptic to the equatorial plane.  However, we will show in Sec. \ref{modLap} that by inducing a slower secular change in $\Omega$, radiation pressure shifts this transition location.  Again using the Uranian system as an example, Fig. \ref{rad} shows the numerical integration of a $Z = 0.1$ particle started at $80 R_p$ with a small seed eccentricity and inclination to the ecliptic.  The inclination follows a modified Laplace plane (cf. Fig. \ref{circ}) as P-R drag slowly decreases the semimajor axis, and $\omega_E$ only becomes stationary when this transition occurs at $a \approx 54 R_p$.  

\subsection{Secular Precession Rates} \label{secprecrates}

\begin{figure}[!ht]
\includegraphics[width=12cm]{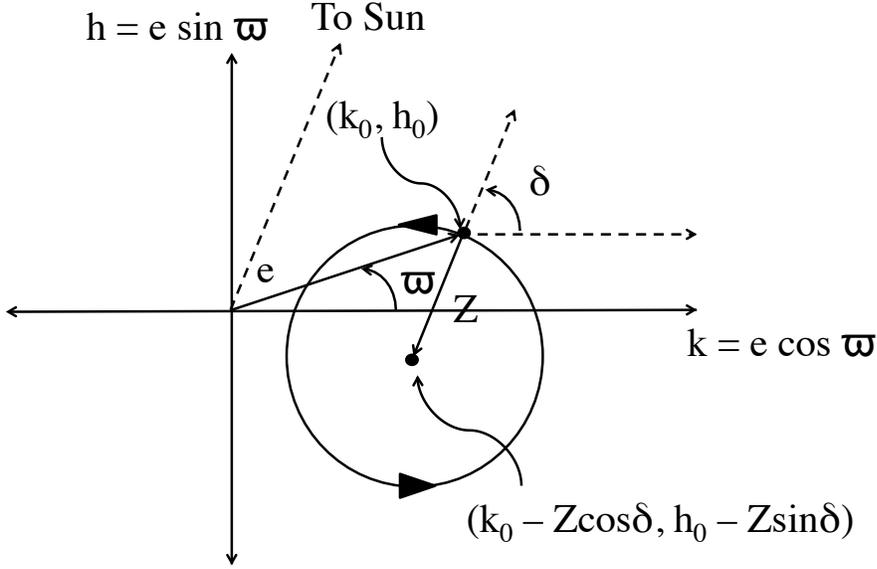}
\caption{\label{pq} Geometrical representation of Eqs. \ref{solpq}.  The system begins at $(k_0, h_0)$ and evolves along the perimeter of the circle of radius $Z$ at a constant rate $n_\odot$.  The orbit's eccentricity $e$ and $\varpi$ can be read as the system point's distance from the origin and polar angle, respectively.  The circle's center relative to $(k_0, h_0)$ is set by the Sun's initial position, $\delta$, and lies at the point $(k_0 - Z\cos \delta, h_0 - Z\sin\delta)$.  }
\end{figure}

Although the equations of motion become more difficult to solve with each added perturbation, we can make some analytic progress for small eccentricities and inclinations to the ecliptic.  The relevant equations of motion, subject to the simplifications mentioned at the beginning of this section, are provided by \cite{Hamilton93}.  The elements are referenced to the ecliptic plane (as there is no ambiguity, we henceforth omit the 'E' subscripts), and since we limit ourselves to low inclinations (ignoring terms of order $i^2$), we switch from $\omega$ to the variable $\varpi = \Omega + \omega$.  This yields
\begin{eqnarray} \label{ham93}
\dot{e} &=& - n_\odot Z \sqrt{1 - e^2} \sin(n_\odot t + \delta - \varpi), \nonumber \\
\dot{\varpi} &=& \frac{n_\odot Z \sqrt{1 - e^2}}{e} \cos(n_\odot t + \delta - \varpi), \nonumber \\
\dot{\Omega} &=& - \frac{n_\odot Z e}{\sqrt{1 - e^2}} \sin (\varpi - \Omega) \sin (n_\odot t + \delta - \Omega),
\end{eqnarray}
where $\delta$ is the angular location of the Sun at $t = 0$ relative to the inertial reference direction.  We approach this system of coupled differential equations through the method of successive approximations.  Expanding Eqs. \ref{ham93} in powers of $e$, we begin by ignoring the terms of order $e$ and higher.  In this limit, $\Omega$ is constant, and the solution for the first two equations is given by \cite{Burns79} in terms of the new variables $k = e\cos \varpi$ and $h = e\sin \varpi$:
\begin{eqnarray} \label{solpq}
k &=& k_0 - Z\cos(\delta) + Z \cos(n_\odot t + \delta) \nonumber \\
h &=& h_0 - Z\sin(\delta) + Z \sin(n_\odot t + \delta),
\end{eqnarray}
where in their solution, the time $t=0$ has been redefined so that $\delta = 0$.  These solutions have a readily-visualized geometric interpretation.  The system evolves at a rate $n_\odot$ along the locus of points defined by a circle of radius $Z$.  The center of this circle is offset from $(k_0, h_0)$ away from the Sun's initial position $\delta$ (see Fig. \ref{pq}).  One can visualize the evolution of $e$ and $\varpi$ from such a representation since the orbital eccentricity at a particular point in $h,\,k$ space is given by the distance from the origin, and $\varpi$ by the polar angle.  The Sun's initial location therefore determines the range of eccentricities explored by setting the location of the circle's center.  	

We now refine our solution by including terms of order $e$ in an expansion of Eqs. \ref{ham93} in powers of $e$.  Omitting the first equation, which is unchanged, 
\begin{eqnarray} \label{2ndorder}
\dot{\varpi} &=& \dot{\varpi_0} - \frac{1}{2}n_\odot Z e \cos(n_\odot t + \delta - \varpi) \nonumber \\
\dot{\Omega} &=& - n_\odot Z e \sin (\varpi - \Omega) \sin (n_\odot t + \delta - \Omega),
\end{eqnarray}
where $\dot{\varpi_0}$ is the zeroth-order rate employed in our first solution.  Expanding the trigonometric functions in the equation for $\dot{\Omega}$, and using the substitutions $k = e \cos \varpi$ and $h = e \sin \varpi$, one obtains,
\begin{equation}
\dot{\Omega} = -n_\odot Z (h \cos \Omega - k \sin \Omega)[\sin (n_\odot t + \delta) \cos \Omega - \cos(n_\odot t + \delta) \sin \Omega].
\end{equation}
We now feed our zeroth order solution back into the above equation.  In particular, we treat $\Omega$ as constant, and use Eqs. \ref{solpq} for $h$ and $k$.  Since we are interested in how radiation pressure interacts with solar tides and oblateness on long secular timescales, we additionally average over a solar cycle from $t=0$ to $t = 2\pi / n_\odot$.  This then yields the simple expression
\begin{equation} \label {Om}
<\dot{\Omega}> \approx - \frac{n_\odot Z^2}{2}.
\end{equation}
For small values of $Z$, this expression is consistent with our previous assumption that $\Omega$ evolves at a rate much slower than $n_\odot$ and matches numerical integrations well.  As $Z$ approaches unity, our approximations worsen.

Applying the same procedure of inserting the zeroth-order solution and averaging over a solar cycle in the expression for $\dot{\varpi}$ in Eq. \ref{2ndorder} yields the same value
\begin{equation} \label{varpi}
<\dot{\varpi}> \approx - \frac{n_\odot Z^2}{2}.
\end{equation}
Since $\varpi = \Omega + \omega$, this means that the secular change in $\varpi$ is entirely due to $\Omega$.  Therefore, to our level of approximation, $\omega$ does not move secularly.  This justifies our claim from Sec. \ref{radpsec} that Eq. \ref{cond} still represents an averaged condition for the halting of pericenter with radiation pressure included.  However, as we argue in the next section, radiation pressure can change the semimajor axis at which the eccentricity becomes unstable by modifying the Laplace surface.

\subsection{The Modified Laplace Surface} \label{modLap}

We showed in the previous section that, to first order in $e$, $\omega$ does not move secularly; therefore Eq. \ref{cond} still holds as a condition for the eccentricity to grow to large values.  However, the regression of $\Omega$ in the ecliptic plane induced by radiation pressure (Eq. \ref{Om}) spoils the Laplace equilibrium between solar tides and planetary oblateness given by Eq. \ref{trans}.  Radiation pressure creates a modified Laplace surface (on which the torques from all three perturbations balance) and shifts the location where the local equilibrium plane transitions from the ecliptic to the equatorial plane.  

For prograde particles, and to first order in $e$ and $i$, the secular regression of the node due to solar tides is given by $\dot{\Omega} = -(3/4) \epsilon_\odot n$ \citep[e.g.,][]{Carruba02Err}.  Radiation pressure thus enhances the nodal regression induced by the Sun.  As a result, the semimajor axis at which these torques balance those from the planet's oblateness (i.e., the point at which the Laplace plane shifts) must move inward, where the effects of the zonal harmonics are stronger.  

We do not calculate the detailed warp of the Laplace plane in this paper---for references on the process in the classical case of the competition between solar gravity and planetary oblateness, see \cite{Ward81} and \cite{Dobrovolskis93}; \cite{Allan64} treat the general case of an arbitrary number of non-interacting perturbers in independent planes; for the different case of radiation pressure offsetting planetary oblateness, see \cite[][cf. Fig. 18 of \citealt{Burns01}]{Hamilton96Mars}.  In our case involving all three perturbations, we limit ourselves to estimating the transition location of the Laplace plane where we expect the orbit to become unstable and execute large-amplitude eccentricity oscillations.  More concretely, this will approximately correspond to the distance at which the nodal precession rates from the solar tides and radiation pressure balance that due to planetary oblateness: $\dot{\Omega}_{Sun} + \dot{\Omega}_{Rad} = \dot{\Omega}_{J2}$.  One can understand this as an approximate condition that the torques from forces directed out of the orbital plane cancel \cite[see Eq. 38 in][]{Burns76}.  The nodal rate due to $J_2$, now referenced to the planet's $\it{equatorial}$ plane and again to first order in the inclination, is $\dot{\Omega}_{J2} = -(3/2) \epsilon_p n$ \citep[e.g.,][p. 270]{Murray99}.  The nodal rate due to solar gravity is $\dot{\Omega}_{Sun} = -(3/4) \epsilon_\odot n$ \citep[e.g.,][]{Carruba02Err}.

Substituting for $\epsilon_\odot$ and $\epsilon_p$ from Eqs. \ref{es} and \ref{ep}, the condition for the balance of precession rates is

\begin{equation} \label{transition}
\frac{3}{4}\frac{M_{\odot} a_T^3}{M_p a_p^3} + \frac{1}{2}Z(a_T)^2\Bigg(\frac{M_\odot a_T^3}{M_p a_p^3}\Bigg)^{1/2} = \frac{3}{2}J_2 \frac{R_p^2}{a_T^2},
\end{equation}
where $a_T$ is the approximate semimajor axis at which the Laplace plane transitions.  For $Z=0$, the equation can be solved analytically, yielding $a_T = 2^{1/5} r_L$.  At Uranus, this corresponds to $74.7 R_p$, which can be seen from Fig. \ref{circ} to be approximately the location where the inclination is intermediate between 0 and Uranus' obliquity of $98^{\circ}$.  It furthermore quite accurately matches the location at which the eccentricity grows rapidly, and is therefore a slightly more accurate estimator than the dimensionally obtained $r_L$. 

\begin{figure}[!ht]
\includegraphics[width=12cm]{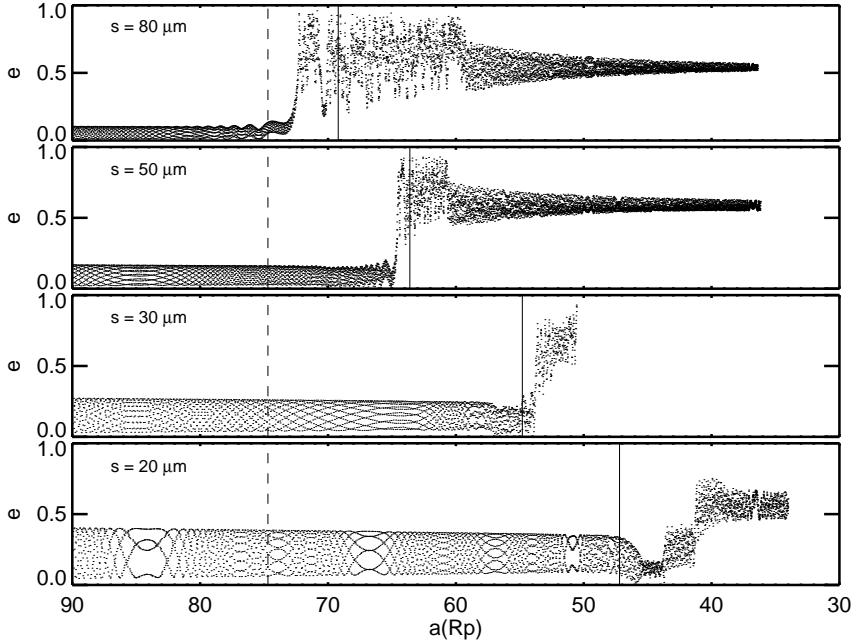}
\caption{\label{zs}  Orbital integrations of particles with various radii $s$ orbiting Uranus.  The four panels, from top to bottom, correspond to values of $Z$ (at $a = 75 R_p$) of 0.04, 0.07, 0.11, and 0.17.  Particles were started at $a=90R_p$ with a seed eccentricity and inclination of $e=10^{-6}$ and $i = 0.06^{\circ}$, respectively.  The vertical solid lines denote the transition locations of the Laplace plane for each size computed from Eq. \ref{transition}.  The dashed lines denote the transition location in the absence of radiation pressure.  This predicted position matches the location where the eccentricity destabilizes to within $\approx 10\%$.}
\end{figure}

One can roughly estimate the particle size-range in which radiation pressure is important at the Laplace plane transition by setting $\dot{\Omega}_{rad} \sim \dot{\Omega}_{Sun}$, or $n_\odot Z(r_L)^2 \sim \epsilon_\odot n(r_L) = n_\odot^2/n(r_L)$, where $r_L$ is the Laplace radius from Eq. \ref{rl}.  This will generally yield a small value of $Z$ since $Z = 1$ would correspond to a Laplace radius equal roughly to the Hill radius.  In the Uranian example previously discussed, this corresponds to $Z \sim 0.05$, or for particles with a density of $1 \text{g}/\text{cm}^3$, to $s \sim 70 \mu$m.  Figure \ref{zs} shows numerical integrations of particles around Uranus with radii $s = 20\mu$m-$80\mu$m, with the corresponding $a_T$ (numerically obtained) marked as a vertical line.

The modified location of the Laplace-plane's transition from Eq. \ref{transition} matches the onset of instability to within $\approx 10\%$.  As the particle size decreases and $Z$ increases, our approximations worsen, and one can see in the fourth panel that the behavior is beginning to change; the eccentricity first decreases, and later rises in two steps.  Our results should therefore be applied with caution beyond $Z \gtrsim 0.2$.  We find that for large values of $Z$, some particles retain low orbital eccentricities as they traverse the unstable region.  We note, however, that this range in $Z$ represents a narrow size range since $Z \propto s^{-1}$.  In this example, the range $Z=0.2-1$ only corresponds to $s\approx20\mu$m-$4\mu$m (particles with $Z \gtrsim 1$ need not be considered as they would have been immediately removed).  If interested in these smallest particles, one must carry out suites of numerical integrations over a wide range of initial conditions to capture the full dynamics.

\subsection{Retrograde Orbits}
We now briefly consider retrograde orbits, which interestingly can exhibit qualitatively different behavior.  For retrograde orbits, $\dot{\Omega}_{Sun}$ and $\dot{\Omega}_{J2}$, which both contain a $\cos i$ dependence, switch sign.  One can obtain $\dot{\Omega}_{rad}$ by rederiving the results of Sec. \ref{secprecrates} starting from the equations given by \cite{Hamilton93} with $i \approx 180^{\circ}$ instead of Eqs. \ref{ham93}; alternatively, one can employ a symmetry argument similar to ones presented in \cite{Hamilton94b}.

One can change a retrograde orbit into a prograde orbit by rotating the coordinate system by $180^{\circ}$ around the $\hat{x}$ axis, so that $\hat{z} \rightarrow - \hat{z}$.  One can then immediately write down the solution found above for prograde orbits, except in this coordinate system the Sun now moves retrograde, so one must make the transformation $n_\odot \rightarrow -n_\odot$.  This yields $\dot{\Omega}^-_{rad} = + n_\odot Z^2 / 2$, where the superscript minus sign denotes that these are elements in the $-\hat{z}$ coordinate system.  The final step is to relate $\Omega^-$ to $\Omega^+$, the longitude of the ascending node in the original coordinate system.  Since, by the right-hand-rule, the directions in which angles increase in the $+\hat{z}$ and $-\hat{z}$ coordinate systems are opposite in direction, $\Omega^+ = - \Omega^-$.  That actually is not quite right, since upon flipping the conventional ``up" direction, the ascending and descending nodes switched places, so $\Omega^+ = 180 - \Omega^-$.  This yields $\dot{\Omega}^+ = - \dot{\Omega}^- = - n_\odot Z^2 /2$; therefore, while the rates due to solar tides and planetary oblateness flip sign for retrograde orbits, the rate due to radiation pressure does not.  This is because, while planetary oblateness and solar tides (after averaging over a solar orbit and smearing the Sun's mass into a ring) are symmetric under $\hat{z} \rightarrow -\hat{z}$, radiation pressure is not, due to the Sun's motion changing sense.

\begin{figure}[!ht]
\includegraphics[width=12cm]{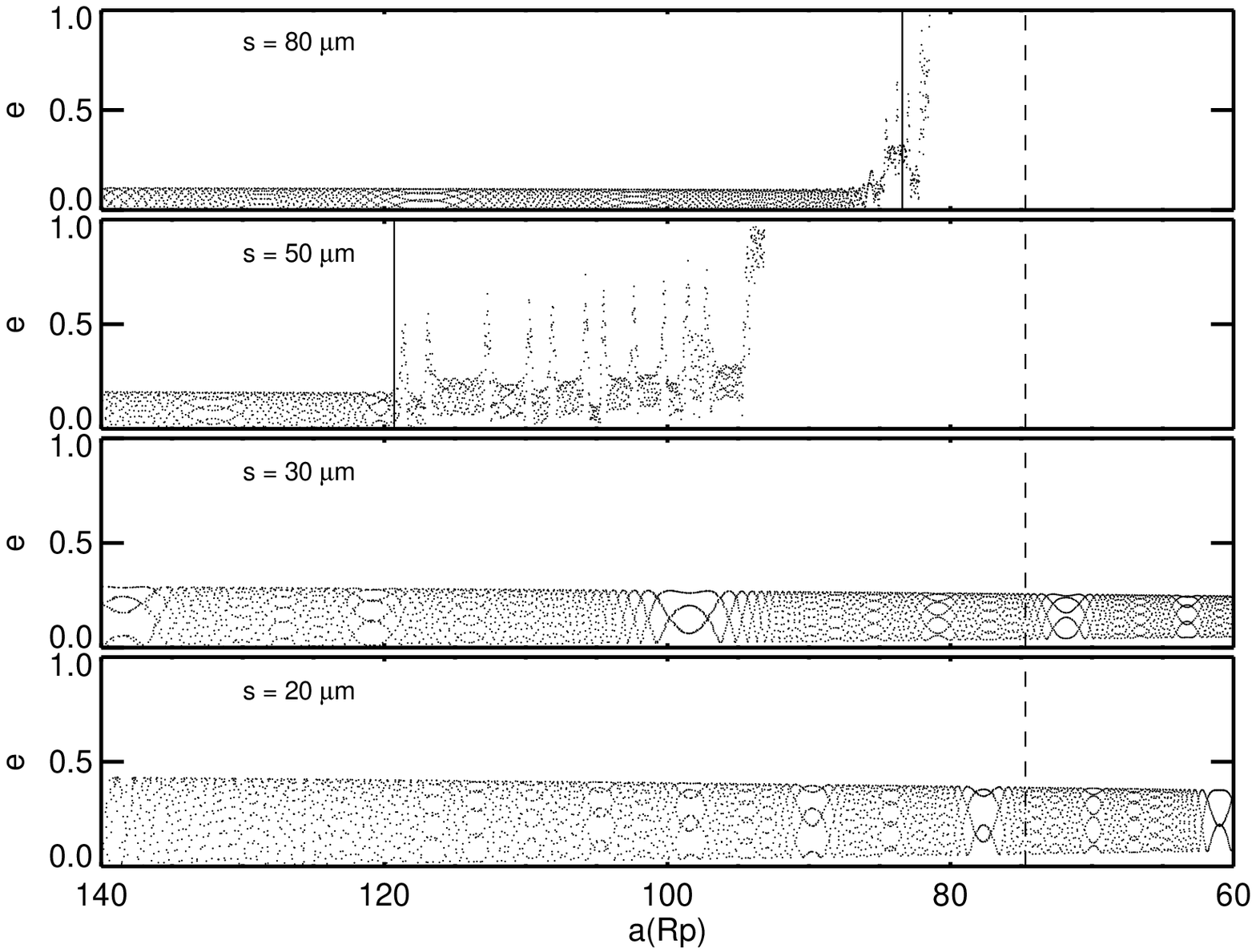}
\caption{\label{i0ZCompr}  Orbital integrations of retrograde particles with various radii $s$ orbiting Uranus.  The four panels, from top to bottom, correspond to values of $Z$ (at $a = 75 R_p$) of 0.04, 0.07, 0.11, and 0.17.  Particles were started at $a=140R_p$ with a seed eccentricity and inclination of $e=10^{-6}$ and $i = 179.91^{\circ}$, respectively.  The vertical solid lines denote the transition locations of the Laplace plane for each size computed from the appropriate condition for retrograde orbits discussed in the text.  The dashed line denotes the transition location in the absence of radiation pressure.  For the bottom two panels, the transition locations are at $a=751 R_p$ and $a = 3803 R_p$, the latter of which is beyond the Hill sphere.  In these two cases, the Laplace plane does not transition to the equatorial plane and the eccentricities remain stable.}
\end{figure}

The condition for the three torques to balance therefore becomes $\lvert \dot{\Omega}_{Sun} \rvert - \lvert \dot{\Omega}_{rad} \rvert = \lvert \dot{\Omega}_{J2} \rvert$.  In this case radiation pressure {\it detracts} from the solar rate, so the transition location will move outward, where weaker oblateness perturbations are sufficient to offset the reduced combination.  There is the further possibility that $\lvert \dot{\Omega}_{rad} \rvert$ overwhelms $\lvert \dot{\Omega}_{Sun} \rvert$, in which case the balance condition cannot be satisfied.  Since $\lvert \dot{\Omega}_{Sun} \rvert \propto a^{3/2}$ while $\lvert \dot{\Omega}_{rad} \rvert \propto a$, there will always exist an $a$ at which the solar rate overtakes the rate due to radiation pressure; however, if that $a$ lies beyond the particle's initial semimajor axis (which is constrained to be smaller than the Hill radius), radiation pressure will always dominate.  In this case there is no Laplace equilibrium and the inclination does not transition to the equatorial plane.  The instability is thereby avoided.  The threshold $Z$ where this occurs is given by the condition $\lvert \dot{\Omega}_{rad} \rvert = \lvert \dot{\Omega}_{Sun} \rvert$.  We considered the balance of these two rates at the Laplace radius in the prograde case; the result evaluated at $a_0$ yields the threshold value of $Z$, $Z_t \approx [3n_\odot/(2n)]^{1/2}$.  Using Eq. \ref{Z} to solve for the threshold particle size $s_t$,
\begin{equation} \label{thresh}
\frac{s_t}{1\:\mu\text{m}} \approx 0.70 Q_{pr} \Bigg(\frac{1\text{ g cm}^{-3}}{\rho}\Bigg) \Bigg(\frac{M_\odot}{M_p}\Bigg)^{1/4} \Bigg(\frac{a_p}{a}\Bigg)^{1/4}.
\end{equation}
The threshold size in the Uranian case with $a_0=140R_p$ is $s_t \approx 46 \mu$m.  Fig. \ref{i0ZCompr} shows the range of behaviors discussed in the previous paragraph for the Uranian case with the same particle sizes as in Fig. \ref{zs}.  The direct integrations match our analytic predictions well for our low chosen values of $Z$.  The irregular behavior in the $50 \mu$m case is presumably the result of its proximity to the threshold size from Eq. \ref{thresh}, but we do not investigate this further in this paper.

\section{CONCLUSION}
We have shown that the unstable range in semimajor axis around planets with high obliquities found by \cite{Tremaine09} can be understood as a modification of Kozai oscillations.  Furthermore, we extended their work (which focused on orbits lying in the local Laplace plane) and provided equations valid for arbitrary inclination.  Although it is difficult to give precise general results, we showed that orbits with inclinations off the Laplace plane are less stable.  We therefore argued that the threshold obliquity of $68.875^{\circ}$ found by \cite{Tremaine09} is an upper limit---inclined orbits can become unstable around planets with lower obliquities.

We then investigated the instability as it applies to dust grains.  Dust grains are subject to Poynting-Robertson drag, which provides a natural mechanism to sweep the semimajor axis inward toward the unstable region.  However, one must also consider the additional effects of radiation pressure on dust-particle orbits.  We found that radiation pressure modifies the classical Laplace surface, and that this shifts the unstable range of semimajor axis.  For prograde particles, this chaotic region is shifted inward, while for retrograde particles it is shifted outward, and can even disappear for small particles.  We estimated the threshold grain size at which orbital eccentricities remain stable for retrograde particles in Eq. \ref{thresh}.  For the smallest particles with $Z \gtrsim 0.2$ (cf. Eq. \ref{Z}), or particles with large initial inclinations or eccentricities, our analytical approximations break down.  We found in simulations that in such cases, for a minority of initial conditions, even prograde orbits can remain stable.  Suites of numerical simulations spanning the range of initial conditions are therefore required to fully characterize a population of dust evolving in toward a high-obliquity planet.

This work can be applied both in the solar system and beyond.  \cite{Bottke10} have proposed that, at each of the giant planets, a vast supply of dust generated by the irregular satellites once existed.  At least in the case of Saturn, this supply persists today \citep{Verbiscer09}.  Many irregular satellites have inclinations close to the low-eccentricity threshold for Kozai oscillations, $i \approx 39.2^{\circ}$ or $150.8^{\circ}$.  These orbits are very unstable, and dust originating from such objects might undergo large-amplitude eccentricity oscillations even around planets with modest obliquities.  At Uranus, all but the smallest particles will do so, and this might explain the color dichotomies common to the outer four regular satellites observed by \cite{Buratti91}.  \cite{Tamayo12} have started toward such an explanation, which we will pursue elsewhere.  More generally, this instability could be applied to myriad classes of circumstellar binary objects, such as binary KBOs and asteroids.  Finally, having incorporated radiation forces, one could consider debris disks in systems with an interior planet (providing an effective $J_2$) and a highly-inclined companion.

\section{ACKNOWLEDGMENTS}
We are grateful to an anonymous reviewer who pointed us to several relevant papers and greatly strengthened this manuscript.  We would like to thank Matija \'Cuk for pointing us to the work of \cite{Tremaine09}, and we further thank Matthew M. Hedman, Matthew S. Tiscareno and Rebecca A. Harbison for insightful comments and discussions.  This research work was supported by the Cassini project and NASA's Planetary Geology and Geophysics Program.

%\bibliographystyle{/Users/dtamayo/Documents/Research/yahapj}
%\bibliography{/Users/dtamayo/Documents/Research/apj-jour,/Users/dtamayo/Documents/Research/Bib}

\end{document}